\documentstyle[12pt,aasms4]{article}

\begin{document}

\title{ Modeling The X-ray Timing Properties Of Cygnus X-1 As Caused By
Waves Propagating In A Transition Disk}

\author{\bf R. Misra}
\affil{Dept. of Physics \& Astronomy, Northwestern University,Evanston, Illinois - 60208, U.S.A }
\authoremail{ranjeev@yawara.astro.nwu.edu}

\begin{abstract}
We show that waves propagating in a transition disk can explain the short term
temporal behavior of Cygnus X-1. In the transition disk model the 
 spectrum is produced by saturated
Comptonization within the inner region of the accretion disk where
the temperature varies rapidly with radius. Recently, the spectrum from
such a disk has been shown to fit the average broad band spectrum of 
this source better than that predicted by the soft-photon 
Comptonization model. Here,
we consider a simple model where waves are propagating cylindrically symmetrically in the transition disk with a uniform propagation speed ($c_p$). We show
that this model can 
qualitatively explain (a) the variation of the power spectral density (PSD)
with energy, (b) the hard lags as a function of frequency and
(c) the hard lags as a function of energy for various frequencies. Thus
the transition disk model can explain the average spectrum and the
short term temporal behavior of Cygnus X-1.

\end{abstract}


\section{Introduction}

The X-ray emission of the well known Black Hole system Cygnus X-1 shows
two long term spectral states. The system spends most of its time
in the hard state where the spectrum can be approximated as an
power-law with an exponential cutoff around 100 keV. The radiative
mechanism generally invoked to explain this spectrum is unsaturated
Comptonization of soft photons in a hot plasma. A model where the
hot plasma is in the form of an corona on top of a cold accretion disk
(Haardt \& Maraschi 1993; Liang \& Price 1977) has been ruled out
by spectral modeling of the spectrum ( Gierlinski et al. 1997;
Nowak et al. 1999). The hot plasma could then be a hot inner accretion
disk with the soft photons either arising from the cold outer disk 
(Shapiro, Lightman \& Eardley 1976), from a pre-shock flow ( Chakrabarti 1997)
or from internally produced Synchrotron photons as in the ADAF model
(Narayan 1996).

In the transition disk model (Misra \& Melia 1996), the structure
of the disk is obtained using the standard $\alpha$-disk model
of Shakura \& Sunyaev (1973) without the assumption that the
effective optical depth ($\tau_{eff}$) is greater than unity. For high 
accretion rates ($ >\sim 10^{18}$ g sec$^{-1}$), $\tau_{eff}$ turns out ( self-consistently) to
be less than unity for the innermost regions. The disk is then
forced to cool by bremsstrahlung self Comptonization which causes
the temperature to be much hotter than that obtained by assuming
black body emission. The temperature rapidly increase with decreasing
radius in this region. Since the Compton y-parameter is always greater
than unity ( although $\tau_{eff} < 1$) the local spectrum is
a Wien peak. The sum over all Wien peak from the different radii
(and corresponding different temperatures) gives rise to a power-law
which matches the observed spectrum of Cygnus X-1 ( Misra \& Melia 1996).
A similar model has been used to explain AGN spectra 
(Maraschi \& Molendi 1990). Fitting broad-band (2 - 200 keV) spectrum
of Cygnus X-1 revealed that the transition disk model fits the data
better than a simple Comptonization one ( Misra et al. 1998;
Misra, Chitnis \& Melia 1998). In particular an ad hoc additional  Wien
peak  ($kT \approx 50 $keV) feature required in the Comptonization 
fit to the spectrum ( Gierlinski
et al. 1997) is not required for the transition disk model fit.

Another way to differentiate between these theoretical models is
by studying the timing analysis of this source. Recently RXTE observed
Cygnus X-1 and measured the power spectral density (PSD) for various energy bins, the time
lags between energy bins as a function of frequency and energy and
the coherence function (Nowak et al 1999a). These observations indicate that
if the time lag is due to light travel time through the plasma, the
size of the plasma has to be around $10^4 GM/c^2$ (Kazanas, Hua
\& Titarchuk 1997) and the seed photon source has to be isotropic
( Nowak et al. 1999b). To avoid the energy problem associated with
such a large plasma, Bottcher \& Liang (1999), proposed a model where
the variations are due to ``blobs'' of cool, dense matter which spiral
inwards through the hot plasma. However, it is not clear whether these
models agree even qualitatively with all 
the trends in the temporal behavior of Cygnus X-1 as observed by RXTE.

Nowak et al. 1999b showed that the time delays could be due to
linear disturbances propagating through the plasma provided the
speed of propagation is much smaller than $c$ and photons of different
energies are predominately produced in different regions. While such
a scenario is not natural in the framework of a Comptonizing plasma,
the transition disk meets the above requirements. In this letter,
we show that waves propagating in a transition disk at least 
match qualitatively with most of the temporal observations.

\section{Waves in the transition disk}

A complete solution for the steady state disk structure in the framework
of the transition disk model can be obtained using physical parameters
which are the mass accretion rate, the black hole mass and the 
the viscosity parameter $\alpha$ (Misra \& Melia 1996). However,
for simplicity and clarity Misra et al. 1997 used an empirical model
to fit the average photon spectrum of Cygnus X-1. In the empirical
model ( which differs from the transition model only in the parameterization 
form) the temperature profile is assumed to be of the form:
\begin{equation}
 kT \propto  r^n \hbox {J}^m\quad \hbox{\rm keV}\;,
\end{equation}
where $r$ is the radius and $\hbox {J}(r) = 1 - (r_i/r)^{1/2}$. $r_i$ is the radius of the last stable orbit, taken to be $r_i = 3 R_g$, with $R_g \equiv2GM/c^2$. Here $n$ and $m$ are free parameters obtained by fitting the spectrum.
The local radiative mechanism for such disks is saturated Comptonization 
(i.e. a Wien peak) since the local y parameter obtained is greater than
unity (Misra et al. 1998).
Since the gravitational power per unit area 
$F_g \propto  R^{-3} \hbox {J}(R)$, the photon flux at earth can be written
as (Misra et al. 1998)
\begin{equation}
f_E(E) = \int^{r_o}_{r_i} F(E,r) dr \propto E^2 \int^{r_o}_{r_i} (kT)^{-4} \exp(-E/kT)\, 
r^{-2}\; dr\qquad \; , 
\end{equation}
where $E$ is the energy of the photon and $r_o$ is the radius to
which the transition disk extends. Misra et al. (1998) fitted this
empirical model and obtained the best fit temperature profile $T(r)$
with $n \approx -2.5$, $m \approx 1$ and the maximum temperature to
be $\approx 70$ keV. This profile is shown in figure 1.

The temporal behavior of the source is studied using the count rate
as a function of time in a certain energy band. It is useful to define
a response function (Nowak et al. 1999b) for a particular energy band  
\begin{equation}
g (r) = \int^{E_2}_{E_1} F(E,r) A (E) dE
\end{equation}
where $E_2$ and $E_1$ are the limits of the energy band and $A(E)$ is
the energy dependent effective area of the detector. In this paper,
we have used approximate values of $A (E)$ for the RXTE detector.
The normalized $g(r)$ for some of the energy bands used in this
paper are shown in figure 1. The plot indicates where most of the photons
contributing to the channel arise in the disk. We have assumed throughout
this paper that the transition disk extends till $50 r_s$.

Following Nowak et al. (1999b), for simplicity, we shall consider
waves propagating cylindrically symmetrically in the disk with a
uniform propagation speed ($c_p$). Further only waves moving toward a 
sink in the origin are considered. The sink at the origin is represented
as $\rho_s ( r, t) = \delta (r) \rho_s (t)$ in the wave equation. In
this formalism the Fourier transform of the observed light curve in
an energy channel ($s(t)$) becomes ( Nowak et al. 1999b, Morse
\& Feshbach 1953)
\begin{equation}
S (f) = {P_s(f) \over 2 \pi} \int^{r_o}_{r_i} g (r) G_f (2 \pi f , r) dr
\end{equation}
where $P_s(f)$ is the Fourier transform of $\rho_s (t)$ and $G_f$ is the
Fourier transform of the relevant two-dimensional Green's function. This
is solved to be  ( Nowak et al. 1999b, Morse \& Feshbach 1953)
\begin{equation}
 G_f (2 \pi f , r) = i \pi H^{(2)}_0 (k r)\; ,
\end{equation}
where $k^2 \equiv (2 \pi f/c_p)^2$ and $H^{(2)}_0$ is the second
Hankel function of order zero.

Several assumptions have been made in the above derivation. First,
the propagation speed ($c_p$) is assumed to be a constant. Second,
the sink is assumed to be located at the origin instead of $r_i$ and
third the temperature profile ($T(r)$ is assumed not to respond
to the wave propagations. These assumptions highlight the limitation
of the analysis to compare with observations. 

\section{Results}

In figure 2 (a), we show the predicted time lags versus frequency between various
energy bands compared with the observed values taken from
Nowak et al. (1999a).
Here, $c_p = 3000 r_g/ secs = 0.2 c$ for
a $10 M_\odot$ Black Hole which is the only free parameter in this
analysis. Only for those values of frequency where
the coherence function is near unity is considered (i.e $f < 10$ Hz). 
Considering the simplifying assumptions made in this analysis the
results do qualitatively match the observations while there are discrepancies 
at higher frequencies. 

Crary et al. (1998) have used approximately 2000 days of data from
the BATSE to estimate the time lag between 20-50 keV and 50 - 100 keV
bands in the frequency range 0.01 to 0.2 Hz. In figure 2 (b) we
show the predicted time lag for the corresponding energy bands. The
slope of the curve is similar to the observations in this frequency
range. However, the observed time lag is $\approx 2$ higher than
the predicted one. This could mean that the wave propagation speed
is generally slightly smaller than $0.2 c$.

In figure 3, we show the predicted time lags versus energy for
three different frequencies. The data are obtained from Nowak et al.
(1999a). The logarithmic dependence for low energies is obtained
for this model. Moreover, the predicted slope of the curves decreases with
frequency matching the observed trend. Again at  high frequency ($\approx 10$)
Hz there is significant deviation from the observations.

The PSD cannot be predicted in the framework of this simple model
since the Fourier transform of the sink term ($P_s(f)$ in eqn (4))
is unknown. Using the observed PSD we obtain $P_s(f)$ and present
the result in figure 4 (a). Figure 4(a) also shows the predicted
PSD$_w$ if $P_s(f)$ was constant (i.e. white noise). It should be noted
that in this model there is no exponential cutoff in PSD$_w$ below 100 Hz.
It would have been difficult to explain the observed PSD if there had
been such a cutoff. Even though, the observed PSD cannot be directly
inferred, the model does make predictions for the ratio of PSD in
various energy bands since this is independent of $P_s(f)$. In
figure 4 (b) we show two calculated ratio as a function of frequency
and compare it with the observed values. The ratios have been
renormalized to one at a frequency 1 Hz.  For a wide range of frequency
the ratios are close to
unity which shows that the PSD is not sensitive to the energy band. One
also finds that there is a remarkable correspondence of the calculated
ratios  with the observed values 
as a function of frequency and for different energy bands. Note that
for frequencies greater than 10 Hz, the present analysis is not
adequate because of the lack of coherence.

In the framework of this model a transition in the temporal behavior
might take place when $k r_{12} \approx 1$ where $r_{12}$ is the distance
between the maxima of $g(r)$ for the two energy bands under consideration.
This would occur at a frequency $f_{tr} \approx c_p/(2 \pi r_{12}) \approx 477/ r_{12} \approx 15$ Hz for $r_{12} \approx 30 r_g$ (figure 1). It is tempting
to associate $f_{tr}$ with the observed frequency where the coherence function
drops below unity. At higher frequencies than $f_{tr}$ damping or non-linear
effects may be dominant giving rise to this effect.

In summary, waves propagating in a transition disk can qualitatively 
explain most of the temporal properties of Cygnus X-1. In the analysis
only one additional parameter to the steady state disk model is used
to explain the temporal behavior of the system.

There are several observational 
features like the high frequency lag and the ``shelves'' in
the lag versus frequency plot which is yet to be explained. The absence
of such features maybe due to the simplifying assumptions (described in
the end of section 2) used in this analysis. 
The transition disk models are generically radiation pressure
dominated and hence are subject to secular and thermal instabilities
(Misra \& Melia 1996). However, these instabilities may be suppressed
by non local effects like advection or convection. Fast moving radial
waves as required in the above model may be produced in such a 
quasi-stable disk.

\acknowledgments

The author would like to thank Ron Taam for useful discussions.

\clearpage

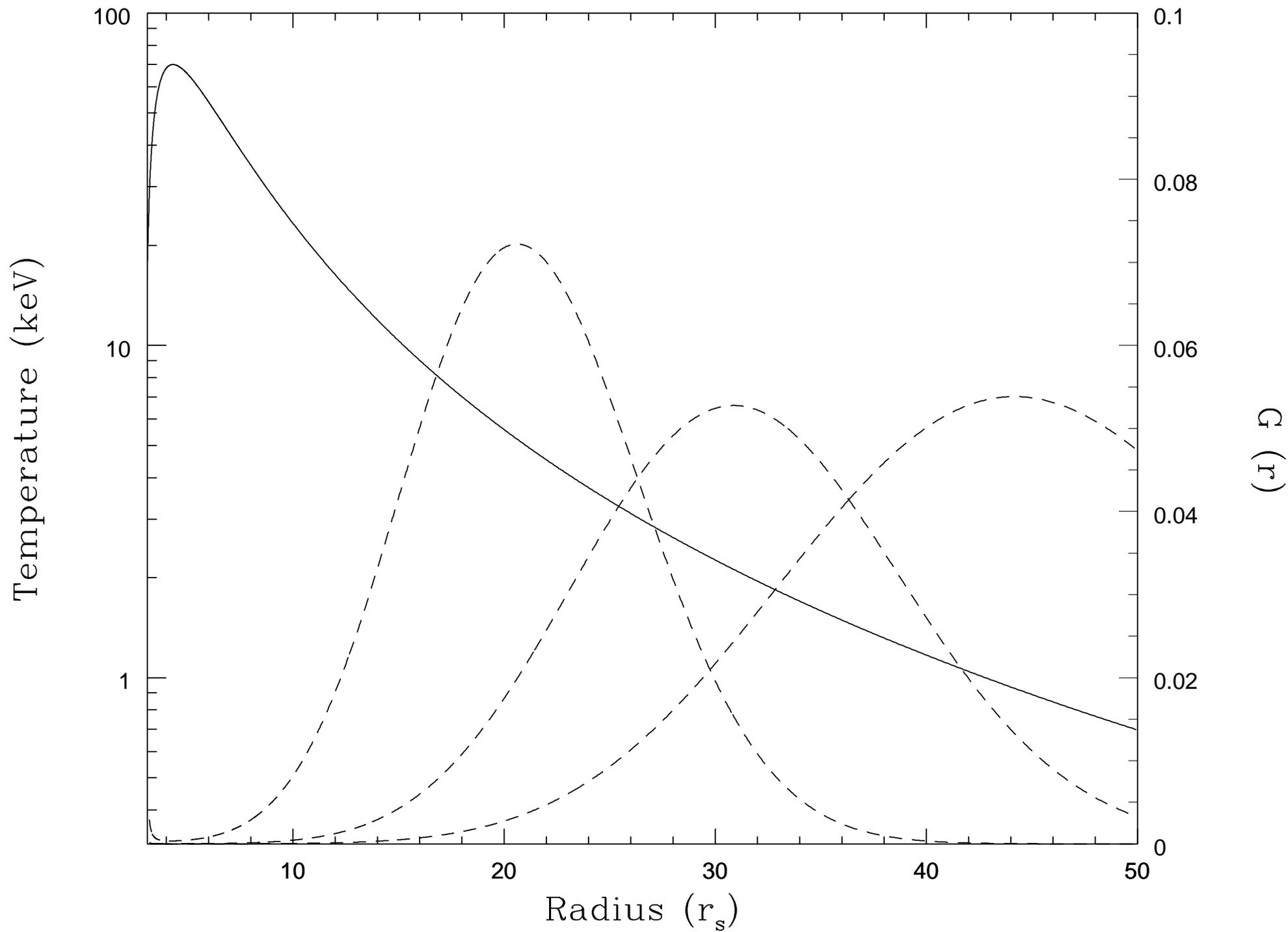
\noindent Fig. 1---Temperature as a function of radius (solid line). The
response function ($g(r)$) for various energy bins (dotted lines).
From left to right the energy bins correspond to (a) 14.1 - 45 keV,
(b) 6.0 - 8.2 keV and (c) 0 - 3.9 keV. 

\noindent Fig. 2 (a)--- Time lags as a function of frequency  between
0 - 3.9 keV and (from top to bottom) (a) 14.1 - 45 keV, (b) 8.2 - 14.1 keV
(solid line) ,(c) 6.0 - 8.2 keV and (d) 3.9 - 6.0 keV. The dots are the observed time
lags taken from Nowak et al. (1999a) between 0 - 3.9 keV and 
8.2 - 14.1 keV.  

\noindent Fig. 2 (b)---Time lags as a function of frequency  between
20 - 50  keV and  50 - 100 keV (solid line). The dashed line
is the best fit to the observed values in the frequency range
0.01 - 0.2 Hz ( Crary et al. 1998). 

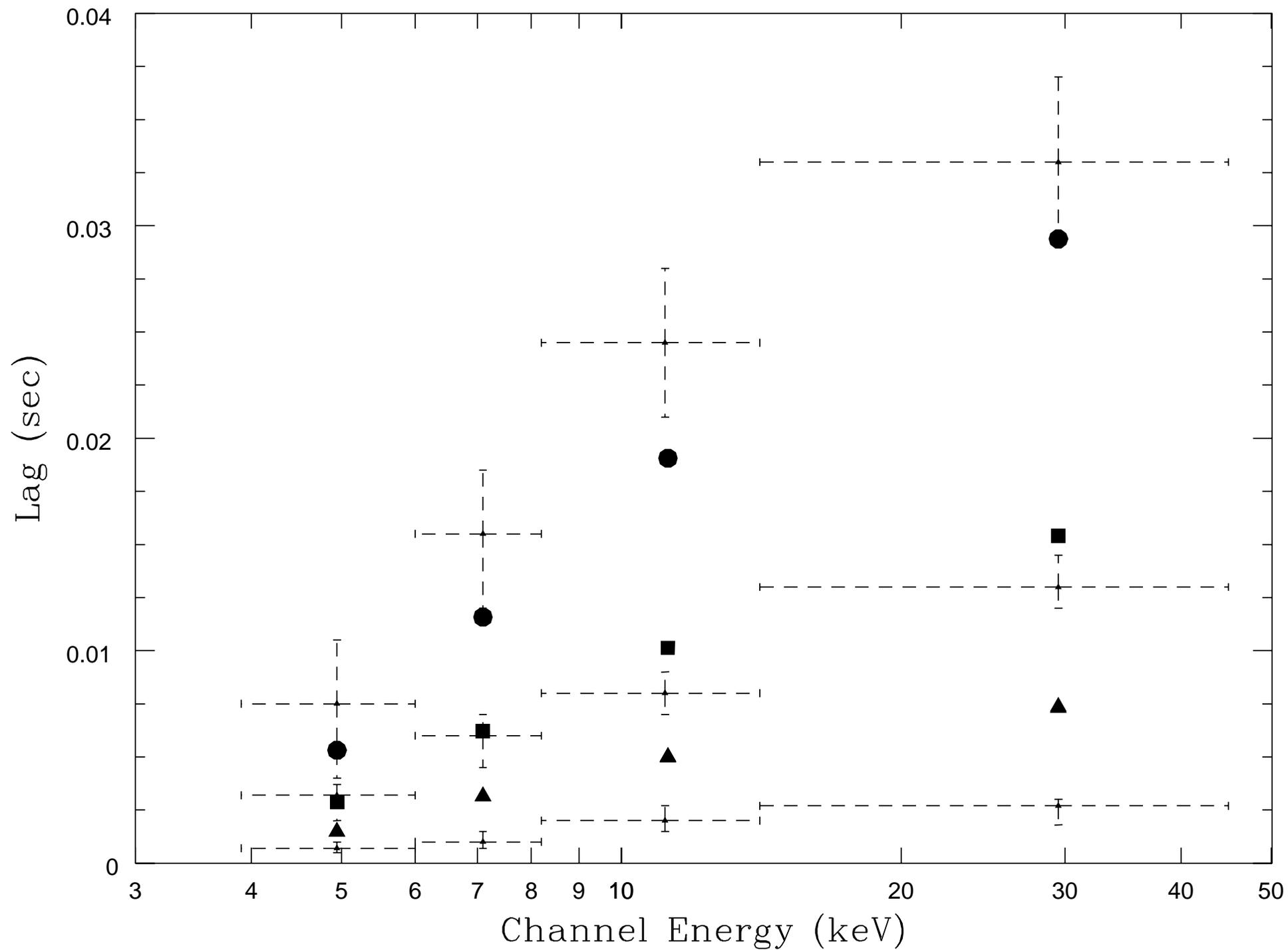
\noindent Fig. 3---Time lag as a function of energy for three different frequencies.
circles: 0.3 Hz, squares: 1 Hz and triangles: 10 Hz. Points with error
bars are data taken from Nowak et al. (1999a) for the corresponding frequencies.

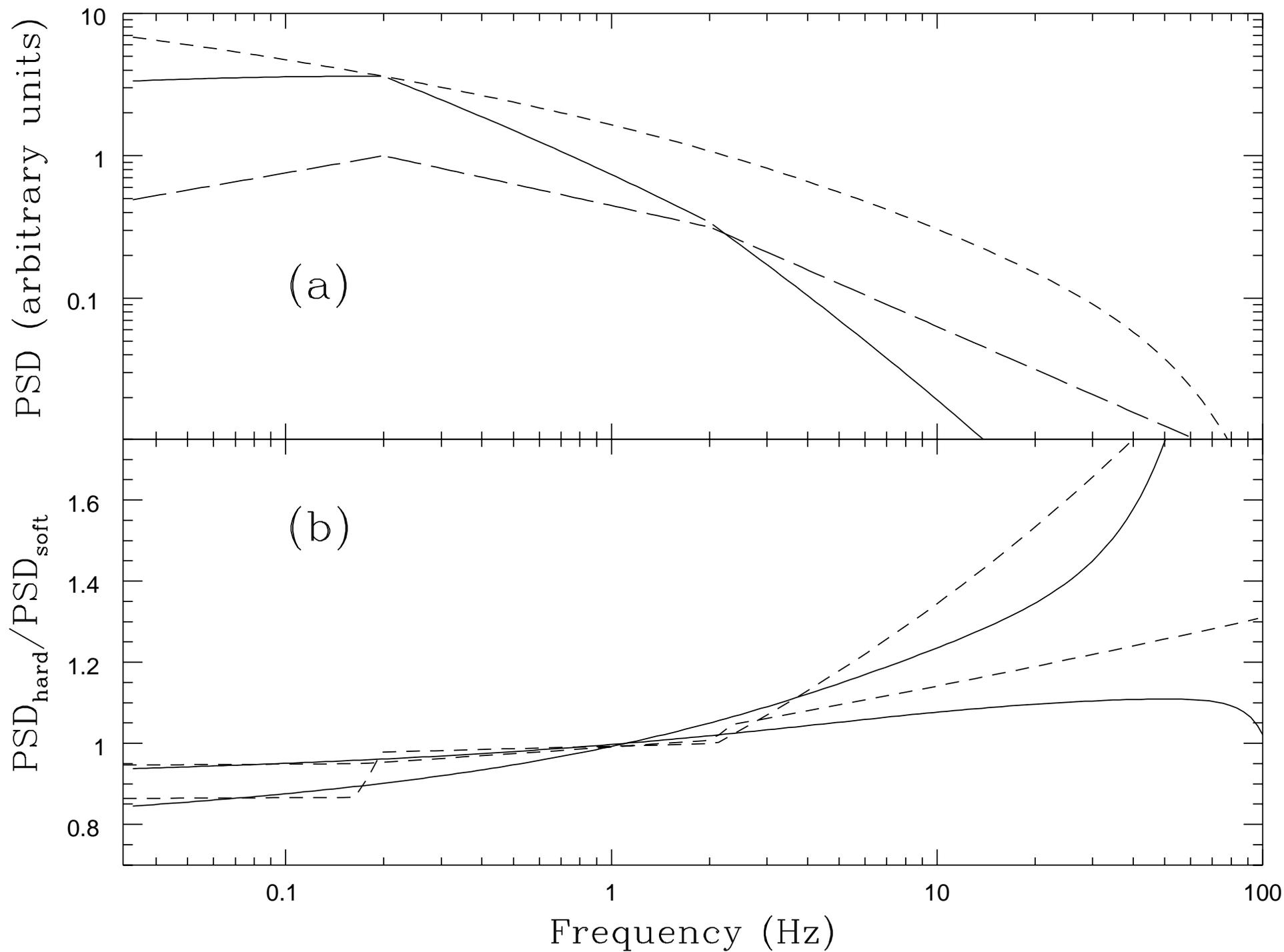
\noindent Fig. 4--- Top:(a) The observed PSD for energy bin 0-3.9 keV (solid line). 
The inferred sink function $ P_S(f)$ required to match the observed PSD (dashed line). The PSD if $P_S(f)$ was a constant (PSD$_w$) which corresponds
to white noise ( dotted line). Bottom (b): The ratio of the PSD. PSD$_{soft}$
is for energy bin 0 - 3.9 keV. PSD$_{hard}$ is for energy bins
14.1 - 45 keV and  6.0 - 8.2 keV. Solid lines are the computed values
while dotted lines correspond to the observed ratios taken from Nowak et al. (1999a).


\begin{references}

\reference{} Bottcher, M., \& Liang , E. P., 1999, \apj , {\bf 511}, L37 Bottcher \& Liang 1999 511 L37

\reference{} Chakrabarti, S.K., 1997, \apj , {\bf 484}, 313.

\reference{} Crary, D. J. et al., 1998, \apj , {\bf 493}, L71.

\reference{} Gierlinski et al., 1997, \mnras, {\bf 288}, 958.

\reference{} Haardt, F. \& Maraschi, L., 1993, \apj, {\bf 413}, 507.

\reference{} Kazanas D., Hua, X. M., \& Titarchuk, L., 1997, \apj, 
{\bf 480}, 280.

\reference{} Liang, E.P. \& Price, R.H., 1977, \apj, {\bf 218}, 247.

\reference{} Maraschi, L. \& Molendi, S., 1990, \apj, {\bf 353}, 452.



\reference{} Misra, R. \& Melia, F., 1996, \apj, {\bf 467}, 405.

\reference{} Misra, R. \& Chitnis, V. R., Melia, F. \& Rao, A.R.  1997, {\bf 487}, 388. 

\reference{} Misra, R. \& Chitnis, V. R. \& Melia, F., 1998, {\bf 495}, 407.

\reference{} Morse, P.M. \& Feshbach, H., 1953, Methods of Theoretical
Physics, Part 1 (New York: McGraw-Hill Book Company, Inc.)

\reference{} Narayan, R. 1996, \apj, {\bf 462}, 136.

\reference{} Nowak, M. A., et al., 1999a, \apj, {\bf 510}, 874.

\reference{} Nowak, M. A., et al., 1999b, \apj, {\bf 515 }, 726.

\reference{} Shakura, N.I. \& Sunyaev, R.A. 1973, {\it Astr. Ap.}, {\bf 24}, 337.

\reference{} Shapiro, S.L., Lightman A.P. \& Eardley, D.M. 1976, \apj, {\bf 204}, 187.


\end{references}
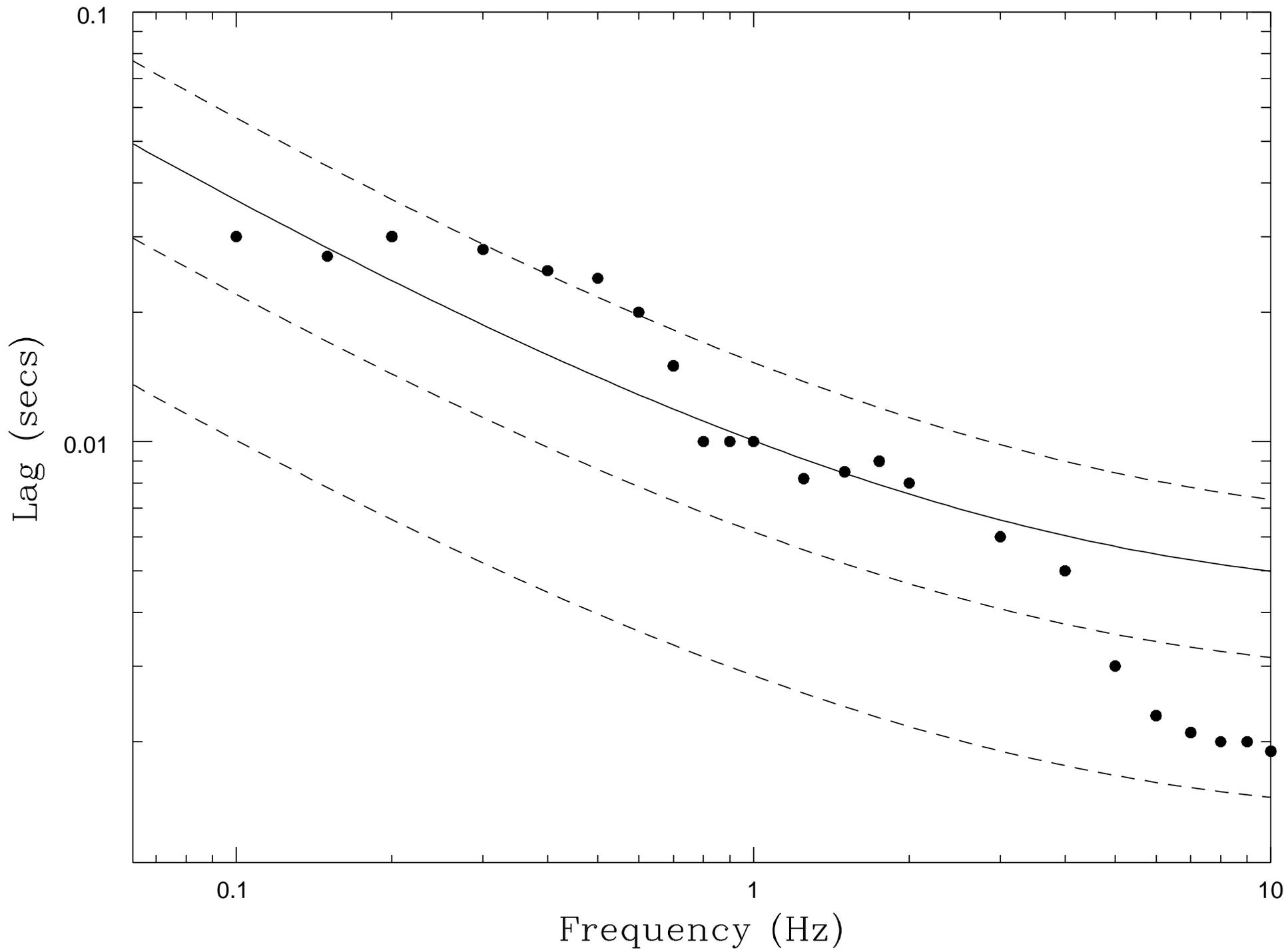
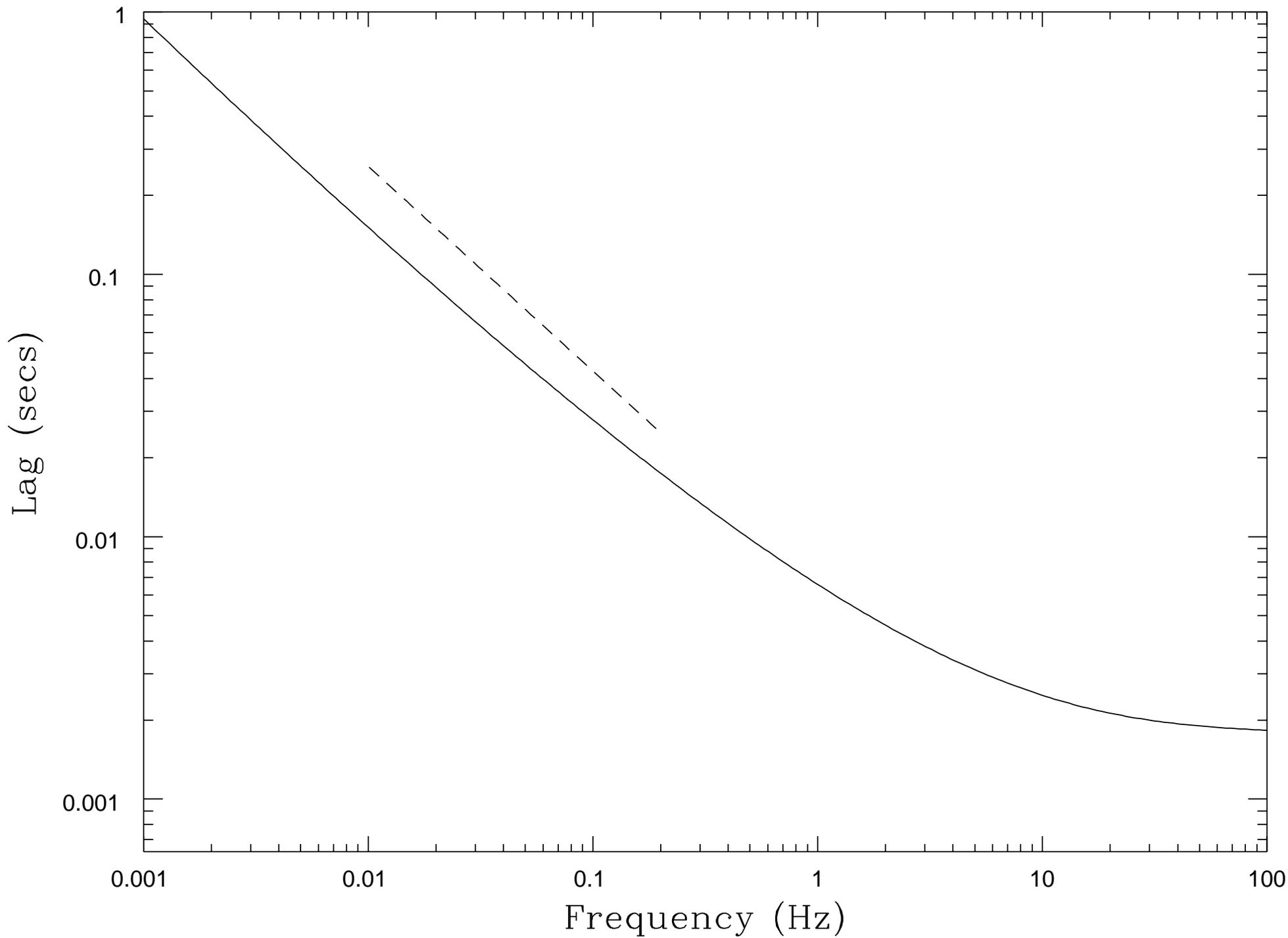
\end{document}